\documentclass[10pt,fleqn,twocolumn,aps,pre,groupedaddress,a4paper,floatfix]{revtex4}

\usepackage{amsmath}
\usepackage{graphicx,psfrag}
\usepackage{multirow}
\usepackage{graphics}
\usepackage{amsfonts}
\usepackage{amssymb}
\usepackage{subfigure}

\begin{document}
 \title{Thermodynamics of quantum Brownian motion with
 internal degrees of freedom:\\ the role of entanglement in the
 strong-coupling quantum regime}
 \author{Christian H\"orhammer}
 \email{christian.hoerhammer@uni-bayreuth.de}
 \affiliation{Theoretische Physik I, Universit\"at Bayreuth, D-95440, Germany}
 \author{Helmut B\"uttner}
 \affiliation{Theoretische Physik I, Universit\"at Bayreuth, D-95440, Germany}
 \date{\today}
 \begin{abstract}
We study the influence of entanglement on the relation between the
statistical entropy of an open quantum system and the heat exchanged
with a low temperature environment. A model of quantum Brownian
motion of the Caldeira-Leggett type -- for which a violation of the
Clausius inequality has been stated by Th.M. Nieuwenhuizen and A.E.
Allahverdyan [Phys. Rev. E 66, 036102 (2002)] -- is reexamined and
the results of the cited work are put into perspective. In order to
address the problem from an information theoretical viewpoint a
model of two coupled Brownian oscillators is formulated that can
also be viewed as a continuum version of a two-qubit system. The
influence of an additional internal coupling parameter on heat and
entropy changes is described and the findings are compared to the
case of a single Brownian particle.
\end{abstract}
\maketitle
\section{Introduction}
Open systems are subject to dissipation of energy and fluctuations
in their degrees of freedom. Within the theory of quantum
dissipative systems \cite{weiss, Dittrich} the starting point in
describing noise and damping is the Hamiltonian $H_{\rm
tot}=H_S+H_E+H_{SE}$ where the Hamiltonian of the total system
$H_{\rm tot}$ is expressed as a sum of the Hamiltonian of the
subsystem of interest $H_S$, a Hamiltonian $H_E$ modeling the
environmental degrees of freedom and an interaction term $H_{SE}$.
For quantum objects this coupling to the environment leads in
addition to the phenomena of decoherence and entanglement. In the
low temperature respectively strong coupling regime these
entanglement effects become important. Under the unitary evolution
of the density operator $\rho_{\rm tot}(t)=U(t,0)\rho_{\rm
tot}(0)U^{\dagger}(t,0)$ with $U(t,0)=\exp[-\frac{i}{\hbar} H_{\rm
tot} t]$ an initial product state of subsystem $S$ and bath $E$
evolves into a correlated state with $\rho_{\rm
tot}(t)\neq\rho_S(t)\otimes\rho_E(t)$ for $t>0$. For zero
temperature the closed, total system is in its ground state and
therefore the von Neumann entropy $S(\rho_{\rm tot}(t))$ stays zero
for all times. But for $t>0$ the pure state of the whole system is
an entangled state of subsystem and bath with $S(\rho_{\rm tot}(t))
\leq S(\rho_S(t))+ S(\rho_E(t))$. The subsystem therefore is in a
mixed state with $S(\rho_S(t))>0$ even for zero bath temperature.

In most applications, especially in quantum optics \cite{gardiner,
breuer}, the coupling between system and bath can be assumed to be
weak which allows neglecting entanglement effects
(Born-Markov-Approximation) and applying the formalism of Markovian
quantum master equations. In this case the statistical
thermodynamics of the open subsystem are governed by the quantum
Gibbs distribution. In the strong-coupling quantum regime, the
stationary state of the subsystem $\rho_S(t)$ is still Gaussian but
non-Gibbsian due to the entanglement with the bath. If the total
Hamiltonian is harmonic, a Gaussian initial state of the subsystem
remains Gaussian for all times. Its density matrix $\rho_S(t)$ is
completely characterized by the first and second moments of the
relevant operators. The Heisenberg equation of motion for these
operators is the quantum Langevin equation \cite{Ford, Ford_Kac}.
The characterizing moments are determined by the stationary values
of the quantum Langevin equation and can be calculated alternatively
by applying the quantum fluctuation-dissipation theorem. The
statistical entropy associated with that stationary quantum state is
the von Neumann entropy. The systems exchange of heat with the
environment -- which is defined as the change in energy due to
redistributions in phase space -- is related to the thermodynamic
entropy by the Clausius inequality. This thermodynamic entropy can
only be identified with the statistical entropy when $\rho_S$ takes
the form of the canonical density matrix. This is just the case for
negligible interaction between subsystem and environment.

From an information-theoretical point of view above considerations
become important. The Landauer principle \cite{landauer, bennett}
which is based on the Clausius inequality states that ``many-to-one"
operations like erasure of information require the dissipation of
energy. Deleting one bit of information is accompanied by a released
amount of heat of at least $kT\ln 2$. This erasure is connected with
a reduction of entropy, and thus cannot be realized in a closed
system. Therefore the information-carrying system has to be coupled
to its environment. To avoid a rapid destruction of the necessary
quantum coherence the quantum subsystem should be placed in a low
temperature environment. Thus the coupling might be relatively
strong compared to thermal energy. Since the Landauer principle is
dealing with information processing and erasure, the relevant
entropy is the statistical entropy of the system. Statistical
entropy and heat are defined separately. Thus,
the relation between both the quantities can be examined.\\
The purpose of our paper is to study deviations from the Clausius
inequality and Landauer bound respectively in the strong coupling
quantum regime. An analytic treatment of this issue is given within
the framework of the Caldeira-Leggett model of quantum Brownian
motion.

In the first part of this paper we want to discuss quite
controversial recent work \cite{Nieuwenhuizen1, Nieuwenhuizen2,
Nieuwenhuizen8, sheehan} and put some of those results into
perspective.  Therefore the quantum Langevin equation of a
harmonically bound quantum particle which is based on the
Caldeira-Leggett model is presented. The stationary moments which
are obtained from this equation characterize the reduced density
matrix which is used to define thermodynamic quantities. Then,
changes in heat and in statistical entropy for adiabatic parameter
variations are compared and the applicability of the Clausius
inequality in the strong coupling/ low temperature quantum regime is
discussed. In the second main part we will focus on a model of
Brownian motion of two coupled oscillators, which can be understood
as a continuum version of a two-qubit system \cite{Rajagopal2}. With
regard to recent work done on continuous variable computing
\cite{Holevo, adesso} the impact on quantum information theory is
studied. It will become clear that additional internal degrees of
freedom can lead to different results as in the case of
quantum motion of a single Brownian particle.\\

\section{Caldeira-Leggett model of quantum Brownian Motion}
Brownian motion is a prominent example of an open quantum system
\cite{haenggi}. The standard model of quantum Brownian motion, often
refereed to as the Caldeira-Leggett model \cite{Caldeira,
Caldeira2}, is a system-plus-reservoir model. The whole system is
governed by the Hamiltonian
\begin{eqnarray}\label{Hcl}H_{\rm tot}&=&\frac{p^2}{2m}+V(q)+{\sum_{i=1}^N\left[{\frac{p_i^2}{2m_i}+\frac{m_i\omega_i^2}{2}x_i^2}\right]}
\nonumber\\&&+{\sum_{i=1}^N\left[-c_ix_iq+\frac{c_i^2}{2m_i\omega_i^2}q^2\right]}\end{eqnarray}
where $q$ and $p$ are the Heisenberg operators for coordinate and
momentum of the Brownian particle moving in the harmonic potential
$V(q)=\frac{1}{2}m\omega_0^2q^2$. The particle is coupled to a bath
of $N$ harmonic oscillators with variables $x_i$ and $p_i$ and
uniformly spaced modes $\omega_i=i\Delta$. The interaction is
bilinear in coordinates of system $q$ and bath $x_i$. For the
coupling-parameters $c_i$ the so-called Drude-Ullersma spectrum
\cite{Ullersma} with large cutoff-frequency $\Gamma$ and coupling
constant $\gamma$ is chosen: $c_i=\sqrt{2\gamma
m_i\omega_i^2\Delta\Gamma^2/\pi(\omega_i^2+\Gamma^2)}$. The bath is
characterized by its spectral density $J(\omega)$, which takes the
form of the Drude spectrum
$J(\omega)=\gamma\omega\Gamma^2/(\omega^2+\Gamma^2)$ in the
thermodynamic limit (sending $\Delta\to 0$, $N\to\infty$ and keeping
$\Gamma=N\Delta=$const.). The potential renormalization term $\sum
c_i^2/(2m_i\omega_i^2)q^2$ ensures that $V(q)$ remains the bare
potential. Neglecting this self-interaction term, the positive
definiteness of the total Hamiltonian $H_{\rm tot}$ would just be
guaranteed for $\gamma\leq m\omega_0^2/\Gamma$ and -- since $\Gamma$
is large -- would restrict the applicability of the model to
weak-coupling approximations ($\gamma\ll m\omega_0$).
\subsection{Quantum Langevin equation}
From the Hamiltonian \eqref{Hcl} the Heisenberg equations of motion
for the operators $q$ and $p$ and the bath variables $x_i$, $p_i$
are received. By eliminating the bath degrees of freedom the quantum
Langevin equation \cite{Ford, Ford_Kac} of a particle moving in the
potential $V(q)$ can be derived:
\begin{equation}\label{langevinBM} m\ddot q(t) + \frac{dV(q)}{dq} +
\int_0^t{dt'\gamma(t-t') \dot q(t')} = \eta(t) -
q(0)\gamma(t).\end{equation} The stochastic character of this
integro-differential equation with the friction kernel
$\gamma(t)=\gamma\Gamma e^{-\Gamma |t|}$ comes into play by
considering the initial distribution of the bath variables which
determines the noise term $\eta(t)$:
\begin{equation}\label{nterm}\eta(t)=\sum_{i=1}^N{c_i\left(x_i(0)\cos\omega_it+\frac{p_i(0)}{m_i\omega_i}\sin\omega_it\right)}.\end{equation}
Assuming an uncorrelated initial state with the reservoir being in
canonical equilibrium at temperature $T=\beta^{-1}$,
$\rho_E\sim\exp(-\beta H_E)$, $\eta(t)$ is a stationary Gaussian
operator noise with $\langle \eta(t)\rangle_{\rho_E}=0$ and the
correlation function \cite{gardiner}:
\begin{eqnarray}\label{kernel}K(t-t')&=&\frac{1}{2}\langle
\eta(t)\eta(t')+\eta(t')\eta(t)\rangle_{\rho_E}=\\&=&\frac{\hbar}{\pi}
\int_0^\infty d\omega \frac{\gamma\Gamma^2\omega}{\Gamma^2+\omega^2}
\coth(\frac{1}{2}\beta\hbar\omega)\cos\omega(t-t').\nonumber\end{eqnarray}
In the case of initial correlations of particle and bath the
correlation function contains additional terms which affect the
dynamics on the timescale $t\leq1/\Gamma$. To fully characterize the
reduced dynamics it is thus important to specify the initial
preparation.

\subsection{Stationary state}
The relaxation dynamics of the moments $\langle q^2(t)\rangle$ and
$\langle p^2(t) \rangle$ described by eq. \eqref{langevinBM}, end up
in a stationary state for $t\to \infty$. The stationary correlations
can be calculated alternatively by applying the quantum
fluctuation-dissipation theorem \cite{Callen}, which establishes a
connection between the quantum mechanical dynamical susceptibility
$\tilde\chi(\omega)=\int_{-\infty}^{\infty}\chi(t-t')e^{i\omega
t}=\left[m\omega_0^2-m\omega^2-i\omega\tilde\gamma(\omega)\right]^{-1}$
and the equilibrium fluctuations $\langle q^2\rangle$ and $\langle
p^2\rangle$:
\begin{equation}\label{xxchi}\langle q^2\rangle=\frac{\hbar}{2\pi}\int_{-\infty}^{\infty}d\omega \coth(\frac{1}{2} \beta \hbar \omega)
\tilde\chi''(\omega),\end{equation}
\begin{equation}\label{ppchi}\langle p^2\rangle=\frac{\hbar}{2\pi}\int_{-\infty}^{\infty}d\omega m^2\omega^2 \coth(\frac{1}{2} \beta \hbar \omega)
\tilde\chi''(\omega).\end{equation} If the dissipative part of the
susceptibility $\tilde\chi''(\omega)$ of the non-Markovian damped
oscillator with three characteristic frequencies $\lambda_1$,
$\lambda_2$ and $\lambda_3$ (poles in the complex plane) is
inserted, an analytic expression for eq. \eqref{xxchi} and
\eqref{ppchi} is derived \cite{weiss}:
\begin{equation}\label{xxeq} \langle q^2 \rangle=\frac{\hbar}{m\pi}\sum_{i=1}^3\frac{(\lambda_i-\Gamma)\psi\bigl(\frac{\beta\hbar \lambda_i}{2\pi}\bigr)}{(\lambda_{i+1}-\lambda_i)(\lambda_{i-1}-\lambda_i)}-T,\end{equation}
\begin{equation}\label{ppeq} \langle p^2 \rangle=m^2 \omega_0^2 \langle q^2 \rangle + \frac{\hbar \gamma \Gamma}{\pi}\sum_{i=1}^3\frac{\lambda_i \psi\bigl(\frac{\beta\hbar \lambda_i}{2\pi}\bigr)}{(\lambda_{i+1}-\lambda_i)(\lambda_{i-1}-\lambda_i)},\end{equation}
where $\psi(x)$ is the Digamma function and $\lambda_0=\lambda_3$,
$\lambda_4=\lambda_1$.  The stationary state of the Brownian
particle is fully characterized by the variances \eqref{xxeq} and
\eqref{ppeq} which determine the stationary density matrix $\rho_S$
of the subsystem \cite{weiss, Grabert}:
\begin{equation}\label{roheq}\rho_S(q,q')=\frac{1}{\sqrt{2\pi \langle q^2 \rangle}} \exp\left[-\frac{(q+q')^2}{8\langle q^2
\rangle}-\frac{(q-q')^2}{2 \hbar^2/\langle p^2
\rangle}\right].\end{equation} This reduced density matrix is
different from the canonical equilibrium density matrix
$\rho_{\rm{th}}\sim\exp(-\beta H_S)$ for any finite coupling
$\gamma$. The statistical entropy of the quantum state $\rho_S$ --
the von Neumann entropy $S(\rho_S)$ -- is \cite{Nieuwenhuizen1,
Agarwal}:
\begin{eqnarray}\label{SvN}
S(\rho_S)&=&-\mbox{Tr}[\rho_S\ln\rho_S]=-\sum_n p_n\ln
p_n=\\&=&(v+\frac{1}{2})\ln(v+\frac{1}{2})-(v-\frac{1}{2})\ln(v-\frac{1}{2})\nonumber,
\end{eqnarray}
with Boltzmann constant set to $k_B=1$ and the subsystems phase
space volume $v$ defined by
\begin{equation}\label{v} v=\frac{1}{\hbar}\sqrt{\langle q^2\rangle \langle p^2\rangle},\end{equation}
as well as the eigenvalues of $\rho_S(q,q')$,
\begin{equation}
p_n=1/(v+1/2)\left[(v-1/2)/(v+1/2)\right]^n,\end{equation} which are
obtained as solution of the problem $\int dx'\rho_S(q,q')f_n(q')=p_n
f_n(q)$, where the eigenfunctions $f_n$ are given by
$f_n\sim\sqrt{c}H_n(cq)e^{-c^2q^2/2}$ with Hermite polynomials $H_n$
and $c=[\langle
p^2\rangle/(\hbar^2\langle q^2\rangle)]^{1/4}$.\\

\begin{figure}[t]
\begin{center}
\psfrag{x1}[c]{$kT/\hbar\omega_0$} \psfrag{x2}[c]{$S(\rho_S)$ and
$S_p$}
 \includegraphics[width=8cm]{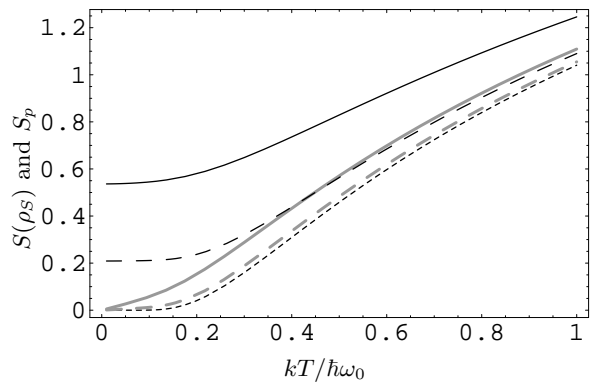}
\caption[von Neumann entropy ]{\label{SvNvsT}\sl\small
  Temperature dependence of the entropy expressions $S(\rho_S)$ and $S_p$ for different values of the
  system-bath couplings $\gamma$ (in units of $m\omega_0^2/\Gamma$).
  Dark lines from bottom to top: $S(\rho_{S})_{\gamma=0};\mbox{ }S(\rho_S)_{\gamma=1};\mbox{ }S(\rho_S)_{\gamma=5}$. Grey lines: $S_{p,\gamma=1}$ (dashed)
  and $S_{p,\gamma=5}$. Other parameters: $\omega_0=1$, $m=1$, $\Gamma=10$,
  $\hbar=1$.}
\end{center}
\end{figure}

The von Neumann entropy of the subsystem is increased with raising
the coupling $\gamma$ at a given temperature (fig. \ref{SvNvsT}).
Even at $T\to0$ the subsystems entropy is larger than zero. This
effect is due to the correlations between subsystem and bath which
prevent the subsystem from reaching a pure state for $T\to0$. The
probabilities to find the subsystem in an exited state depend on the
coupling to the environment \cite{jordan}. In the weak coupling
limit, where $\rho_S=\rho_{\rm th}$, expression \eqref{SvN} gives
the entropy
\begin{equation}\label{Sbose}
S(\omega_0,T)=\frac{\beta\hbar\omega_0}{e^{\beta\hbar\omega_0}-1}-\ln\left(1-e^{-\beta\hbar\omega_0}\right)
\end{equation}
of an harmonic oscillator in canonical equilibirum.\\
It is important to remark here, that the entropy \eqref{SvN}
deviates from the difference of the total entropy $S(\rho_{\rm
tot})$ and the entropy of the bath in absence of the particle
$S(\rho_E)_{\gamma=0}$ which is given by {\cite{Ford2}}
\begin{equation}\label{Sp}
S_p=\frac{1}{\pi}\int_0^{\infty}S(\omega,T)\mbox{Im}\left\{\frac{d\ln\tilde\chi(\omega)}{d\omega}\right\}d\omega.
\end{equation}
In the same way the thermodynamic potentials $F_p$ and $U_p$ can be
derived which are related by $S_p=\beta(U_p-F_p)$. One can see from
fig. \ref{SvNvsT} that $S_p$ vanishes at $T\to 0$ whereas
$S(\rho_S)$ does not. Proceeding in this way, the entropy $S_p$ also
contains the part of entropy that is associated with the quantum
mechanical correlations or particle and bath. Since this conditional
entropy is negative for entangled systems, the statistical entropy
of the Brownian oscillator alone is underestimated by $S_p$. Thus,
in our further treatment we will concentrate on the entropy
$S(\rho_S)$.

\subsection{Thermodynamics of adiabatic changes}

Nieuwenhuizen and Allahverdyan \cite{Nieuwenhuizen8} examined the
validity of the Clausius inequality respectively the Landauer bound
in the strong coupling quantum regime. They found a violation of
these principles at very low temperatures due to the existing
correlations between subsystem and bath. With respect to quantum
information theory they concluded that quantum mechanical
information carrier therefore could be more efficient than their
classical counterparts. A controversial subject in this context is
the appropriate choice of heat and entropy expressions. Our purpose
in this section is to clarify this issue and to put the findings of
Nieuwenhuizen and Allahverdyan into perspective.

The internal energy of the Brownian oscillator can be defined as the
mean energy in the stationary state \cite{Nieuwenhuizen1}:
\begin{equation}
U_S=\mbox{Tr}\left[H_S\rho_S\right]=\langle
H_S\rangle=\frac{1}{2m}\langle
p^2\rangle+\frac{1}{2}m\omega_0^2\langle q^2\rangle.
\end{equation}
This expression differs from the equivalent to \eqref{Sp} defined
internal energy $U_p$. The difference $U_p-U_S$ can be interpreted
as the interaction energy $U_{\rm int}$ ($\neq \langle
H_{SE}\rangle$!) which is related to the free energy $F_p$ by
\begin{equation}
U_{\rm int}=U_p-U_S=\Gamma\frac{\partial F_p}{\partial \Gamma}.
\end{equation}
Choosing the parameter values $m,$ $\omega_0$ and $\Gamma$ as in
fig. \ref{SvNvsT} the ratio $\tau=U_{\rm int}/U_p$ at zero bath
temperature is given by
$\tau_{(\gamma=m\omega_0^2/\Gamma)}\approx0.03$ and
$\tau_{(\gamma=5m\omega_0^2/\Gamma)}\approx0.10$. For
$kT=\hbar\omega_0$ the ratios are
$\tau_{(\gamma=m\omega_0^2/\Gamma)}\approx0.01$ and
$\tau_{(\gamma=5m\omega_0^2/\Gamma)}\approx0.05$ respectively. \\

The total differential $dU_S$ of the internal energy $U_S$,
\begin{equation}\label{dUS}dU_S=\mbox{Tr}[\rho_S dH_S]+\mbox{Tr}[H_S d\rho_S]=\delta W+ \delta Q\end{equation}
can be divided into two parts \cite{balian}. The first term results
from the change of the parameters $m$ and $\omega_0$ in the
Hamiltonian, so it is a mechanical, non-statistical object and will
be referred to as work $\delta W$:
\begin{equation} \delta W = m\omega_0\langle q^2 \rangle d
\omega_0+\left(\frac{\omega_0^2\langle q^2\rangle}{2} -\frac{\langle
p^2 \rangle}{2m^2}\right)dm.
\end{equation}
The second term $\mbox{Tr}[H_Sd\rho_S]$ represents the variation of
$U_S$ due to the statistical redistribution of the phase space,
which will be associated with the change in heat $\delta Q$:
\begin{eqnarray}
\delta Q &=& \delta_{\omega}Q+\delta_{m}Q \qquad
\mbox{with}\nonumber\\
\delta_{\omega}Q&=&\left(\frac{1}{2}m\omega_0^2\frac{\partial\langle
q^2 \rangle}{\partial \omega_0}+\frac{1}{2m}\frac{\partial \langle
p^2 \rangle}{\partial
\omega_0}\right)d\omega_0,\\
\delta_{m}Q&=&\left(\frac{1}{2}m\omega_0^2\frac{\partial \langle q^2
\rangle}{\partial m}+\frac{1}{2m}\frac{\partial \langle p^2
\rangle}{\partial m} \right)dm
\end{eqnarray}
Now the validity of the Clausius inequality
\begin{equation}\label{Clausius}
\delta Q \leq T dS
\end{equation}
can be evaluated. The Second Law of thermodynamics in the
formulation by Clausius states that in a quasi-static process,
during which the system at all times passes through equilibrium
states, one has $dS_{\rm th}=\delta Q/T$. The thermodynamic entropy
$S_{\rm th}$ defined by the Clausius equality can only be identified
with the statistical entropy $S(\rho_S)$ at thermal equilibrium
where $\rho_S=\rho_{\rm th}$ with $\rho_{\rm
th}=Z^{-1}\exp\left(-\beta H_S\right)$ and
$Z=\mbox{Tr}\exp\left(-\beta H_S\right)$, because
\begin{eqnarray}
dS&=&-\mbox{Tr}[d\rho_{\rm th}\ln\rho_{\rm th}]=\mbox{Tr}[d\rho_{\rm
th}\ln
Z]+\beta\mbox{Tr}[d\rho_{\rm th}H_S]=\nonumber\\
&=&\beta\mbox{Tr}[\rho_{\rm th}H_S]=\beta \delta Q=dS_{\rm th}.
\end{eqnarray}
Concerning the Landauer principle, which is based on the Clausius
inequality, but deals with information processing and erasure, the
relevant entropy is the statistical entropy $S(\rho_S)$. Therefore
we want to compare changes in the statistical entropy $dS$ to
changes in heat $\delta Q$ induced by adiabatic variation of the
systems parameters. The total differential of the von Neumann
entropy \eqref{SvN} is given by
\begin{equation}\label{dSrho}
dS=-\mbox{Tr}[d\rho_S\ln\rho_S]=\ln\left(\frac{v+\frac{1}{2}}{v-\frac{1}{2}}\right)dv.
\end{equation}
Thus, the sign of the change in $S(\rho_S)$ is determined by the
sign of the change in $v$. Note here that the parameters $m$ and
$\omega_0$ are chosen as independent quantities, so that we can
examine $\delta Q_m \leq T dS_m$ and $\delta Q_{\omega}\leq
TdS_{\omega}$ separately.
\begin{figure}[h]
\begin{center}
\psfrag{x1}[c]{$kT/\hbar\omega_0$} \psfrag{x2}[c]{$TdS_m$ and
$\delta Q_m$}
 \includegraphics[width=8cm,clip]{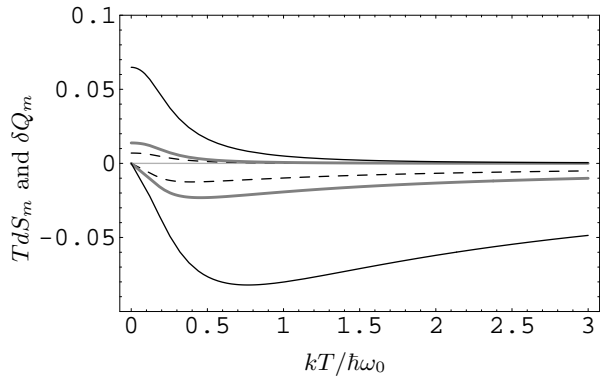}
 \caption[changes in heat and entropy]{\label{dQmdSmvsT}\sl\small
   Changes in heat $\delta Q_m$ (positive) and entropy-term $TdS_m$ (negative) versus bath
   temperature $T$ for different values of the system-bath coupling $\gamma$ (in units of $m\omega_0^2/\Gamma$):
   0.5 (dashed), 1 (grey), 5 (dark). The oscillator parameters are chosen to be $\omega_0=1$ and $m=1$. The cutoff-frequency is set to $\Gamma=10$ and $\hbar=1$.}
 \end{center}
\end{figure}\\
Fig. \ref{dQmdSmvsT} shows the temperature dependence of the changes
in heat $\delta Q_m$ and of the term $TdS_m$ for different coupling
strength $\gamma$. While $\delta Q_m$ is always positive, which
means that the Brownian particle absorbs heat during an adiabatic
increase of mass, the change of entropy and therefore the product
$TdS_m$ remains negative. In the high-temperature limit one has
$\delta Q_m \to 0$ as well as $TdS_m \to 0$ and therefore the
behavior of an uncoupled harmonic oscillator characterized by
\eqref{Sbose}. The term $TdS_m$ converges relatively slowly towards
zero because of the increasing factor $T$. The smaller the coupling
$\gamma$, the faster is the convergence of the two terms $\delta
Q_m$ and $TdS_m$. If the temperature $T$ goes to zero, then the
product $TdS_m$ does as well. The amount of heat $\delta Q_m$
exchanged with the bath stays positive even in this limit and equals
$-d_m U_{\rm int}$. Thus, the Brownian particle can extract heat
from the bath even at $T=0$, a fact that was already extensively
discussed in ref. \cite{Nieuwenhuizen1, Nieuwenhuizen8}.\\
\begin{figure}[ht]
\begin{center}
\psfrag{x1}[c]{$m$} \psfrag{x2}[c]{$v=\sqrt{\langle q^2\rangle
\langle p^2\rangle}/\hbar$}
 \includegraphics[width=8cm,clip]{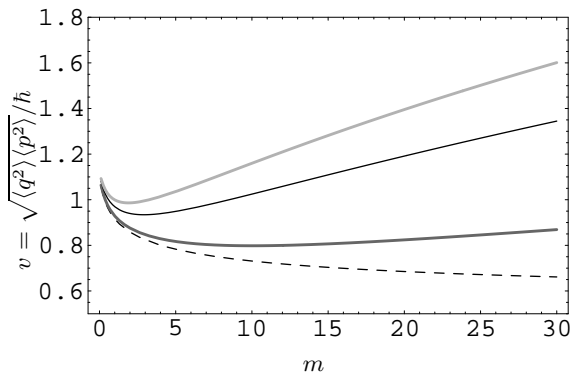}
 \caption[phase space volume]{\label{vniewenh1}\sl\small
   Phase space volume $v$ versus mass m for different $T$-values.
   From bottom to top: $T=0;\mbox{ } 0.1;\mbox{ } 0.2;\mbox{ } 0.25$. Oscillator potential $V(q)=\frac{1}{2}aq^2$ with fixed spring constant $a=1$.
   Other parameters: $\gamma=1$, $\Gamma=500$, $\hbar=1$. (See also ref. \cite{Nieuwenhuizen8}.)}   \end{center}
\end{figure}

Fig. \ref{vniewenh1} gives results of the cited work
\cite{Nieuwenhuizen8}. Nieuwenhuizen and Allahverdyan studied the
influence of adiabatic changes in mass on the phase space volume $v$
given by eq. \eqref{v}. At $T=0$ one receives a monotonous
decreasing function which should converge with increasing mass to
the minimal value $v_{min}=1/2$ which results from the uncertainty
relation. They found, that at moderate temperatures the phase space
volume first decreases for low masses, then reaches a minimum and
finally increases nearly linearly with the mass. The increasing
phase space volume means a positive sign for the entropy change
$dS$. Therefore the authors conclude that the different signs of
$\delta Q$ and $TdS$ would only occur at very low temperatures due
to quantum correlations between system and bath.\\

\begin{figure}[ht]
\begin{center}
\psfrag{x1}[c]{$m$} \psfrag{x2}[c]{$v=\sqrt{\langle q^2\rangle
\langle p^2\rangle}/\hbar$}
 \includegraphics[width=8cm,clip]{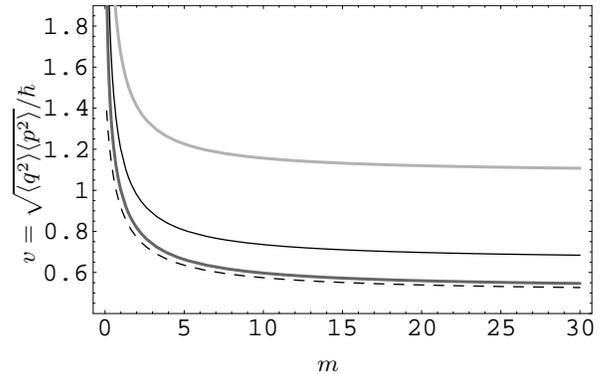}
 \caption[phase space volume]{\label{vmwstatta}\sl\small
  Phase space volume $v$ versus oscillator mass m for different temperature values.
   From bottom to top: $T=0;\mbox{ } 0.25;\mbox{ } 0.5;\mbox{ } 1$. Oscillator potential $V(q)=\frac{1}{2}m\omega_0^2q^2$ with frequency $\omega_0=1$.
   Other parameters: $\gamma=1$, $\Gamma=500$, $\hbar=1$.}\end{center}
\end{figure}

In contrast to that, fig. \ref{vmwstatta} shows the mass dependence
of the phase space volume using the moments defined by eq.
\eqref{xxeq} and \eqref{ppeq}. One can clearly see that even for
moderate temperatures the phase space volume does not increase, but
reaches a temperature-dependent limit value. This value is given by
the phase space volume of an uncoupled harmonic oscillator in
canonical equilibrium:
$v_{th}=\frac{1}{2}\coth{\beta\hbar\omega_0/2}$. Both, sending $m
\to \infty$ or coupling $\gamma \to 0$, finally leads to the
standard case of the quantum Gibbs distribution.\\

The differences between our findings and the results in
\cite{Nieuwenhuizen8} which become obvious in the figures
\ref{vniewenh1} and \ref{vmwstatta} can be explained as follows: in
the paper by Nieuwenhuizen and Allahverdyan \cite{Nieuwenhuizen8}
the harmonic potential $V(q)=\frac{1}{2}m\omega_0^2q^2$ is expressed
by $\frac{1}{2}aq^2$ with spring constant $a$. Varying $m$ and
keeping $a$ fixed leads to the results of fig. \ref{vniewenh1}. But
since the limit $m \to \infty$ and $\gamma \to 0$ should lead to the
same result of $v=1/2$ at T=0, this choice of the potential is
inconsistent. In fig. \ref{vniewenh1} the phase space volume at
$T=0$ does not reach the value $v=1/2$ even for high masses (instead
$v\approx 0.6$ for the given parameter values).  In order to receive
the correct expressions for $\langle q^2\rangle$, $\langle
p^2\rangle$ and $U$ in the weak coupling or high temperature limit
one has to set $a=m\omega_0^2$, which has also been done in the
cited work \cite{Nieuwenhuizen8} in different contexts.\\

Choosing the potential $V(q)=\frac{1}{2}m\omega_0^2q^2$, which is
then affected by the variation of the mass, shows that the anomaly
of different signs is an even stronger effect than found in
\cite{Nieuwenhuizen8}. This indicates that not only quantum
correlations at low temperatures might play a role but also
classical correlations between the damped oscillator and its
environment at moderate temperatures.

\section{Quantum Brownian Motion of two coupled harmonic oscillators}
In order to study the influence of additional parameters on the
results stated above we introduce a model of quantum Brownian motion
of two coupled oscillators which can be viewed as a continuum
version of a two qubit system. In contrast to former work concerning
the relaxation dynamics of two coupled oscillators
\cite{Rajagopal2,Zoubi} we focus on the stationary state. The
Hamiltonian $H_{S'}$ of the open quantum system $S'$ now reads
\begin{equation}\label{Hs}
H_{S'}=H_A+H_B+H_{AB},
\end{equation}
where $H_A$ and $H_B$ are the Hamiltonians of the two harmonic
oscillators $A$ and $B$ with masses $m_a, m_b$ and frequencies
$\omega_a$ and $\omega_b$. $H_{AB}$ describes the interaction
between them. Before deriving a quantum Langevin equation it is
necessary to discuss different couplings between the oscillators as
well as between the system $H_{S'}$ and the bath $H_E$ and to choose
an appropriate model.
\subsection{Coupling between the two oscillators}
In the framework of quantum optics the coupling between oscillators
is often chosen to $H_{AB}=-Dq_aq_b$ with coupling parameter $D$. In
this case one problem is the constraint $D \leq
\sqrt{m_am_b}\omega_a\omega_b$ as a condition for real
eigenfrequencies of the system which restricts the range of allowed
parameter variations. In our further treatment we will concentrate
on the interaction Hamiltonian
\begin{equation}\label{Hac3}H_{AB}=\frac{1}{2}D(q_a-q_b)^2.\end{equation}
This Hamiltonian is clearly inspired by its mechanical analogy
 -- a restoring force proportional to the relative distance of the two oscillators --
and leads to real eigenfrequencies of the system $H_{S'}$ for all
values of the coupling parameter $D$.
\begin{figure}[t]
\begin{center}
\psfrag{x1}[c]{$\gamma/m\Gamma$}\psfrag{x2}[c]{$\mbox{Re}\{\nu_i\}$}
 \includegraphics[width=8cm,clip]{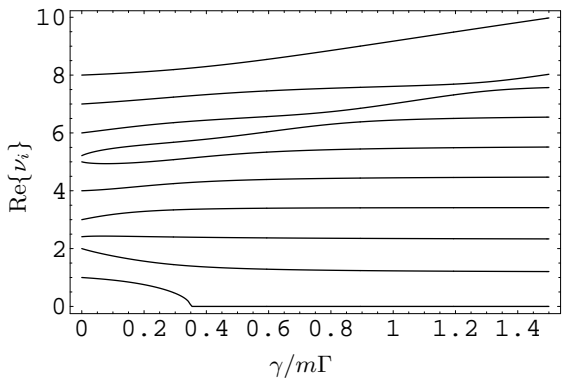}
 \caption[Bildunterschrift]{\label{EWsep}\sl\small
 Oscillators separately coupled to the bath: eigenfrequencies $\nu$ of eq. \eqref{nusep} for a finite model with $N+2=10$ oscillators versus the
 system-bath coupling strength $\gamma$ (in units of $m\Gamma$). The critical value $\gamma{\rm crit}$ is given by eq. \eqref{gammakrit}. Parameters: $\omega_a=2$, $\omega_b=5$, $m=m_{a,b}=1$, $D=2$ and $\Delta=1$, $\Gamma=N\Delta=8$.
  }  \end{center}
\end{figure}
\subsection{Coupling between system and bath}
In order to study the case of strong system-bath-coupling $\gamma$,
the positive definiteness of the Hamiltonian has to be guaranteed in
the range of relevant $\gamma$-values. Therefore we will discuss
different couplings between system and bath in the following
section.
\subsubsection{Oscillators separately coupled to the bath}
If each oscillator is coupled separately to the bath according to
the coupling in \eqref{Hcl} then we receive the following
interaction Hamiltonian $H_{S'E}$:
\begin{equation}
H_{S'E}={\sum_{i=1}^N\left[-c_ix_{i}(q_a+q_b)+\frac{c_i^2}{2m_{i}\omega_{i}^2}(q_a^2+q_b^2)\right]}
\end{equation}
The equation for the eigenvalues $\nu$ of the total system
\begin{equation}H_{\rm tot}=H_A+H_B+H_{AB}+H_E+H_{S'E}\end{equation} reads:
\begin{eqnarray}\label{nusep}\nu^2-\left(\omega_a^2+\frac{D}{m_a}\right)=\sum_i\frac{c_i^2}{m_am_i(\nu^2-\omega_{i}^2)}\frac{\nu^2}{w_{i}^2}-\nonumber\\
-\frac{\frac{1}{m_am_b}\left(D-\sum_i\frac{c_i^2}{m_{i}(\nu^2-\omega_{i}^2)}\right)^2}
{\omega_b^2+\frac{D}{m_b}+\sum_i\frac{c_i^2}{m_bm_i(\nu^2-\omega_{i}^2)}\frac{\nu^2}{\omega_{i}^2}-\nu^2}.\qquad\end{eqnarray}
Fig. \ref{EWsep} shows the influence of the system-bath-coupling
strength $\gamma$ on the eigenfrequencies of a finite system
consisting of the two oscillators coupled to an environment of eight
oscillators. The lowest eigenvalue $\nu_1$ decreases with increasing
coupling strength and becomes imaginary at a critical value
\begin{equation}\label{gammakrit}
\gamma_{\rm
crit}=\frac{\sqrt{(m_a\omega_a^2+D)(m_b\omega_b^2+D)}-D}{\frac{2\pi}{\Delta}\sum_{i=1}^N\frac{\Gamma^2}{\omega_i^2+\Gamma^2}}
\end{equation}
which means exponentially increasing amplitudes and therefore
instability of the whole system. In the thermodynamic limit this
critical value becomes very small, so that this model is only
suitable in the weak coupling case.

\subsubsection{Bath-coupling to the center of mass}
\begin{figure}[t]
\begin{center}
\psfrag{x1}[c]{$\gamma/m\Gamma$}\psfrag{x2}[c]{$\mbox{Re}\{\nu_i\}$}
 \includegraphics[width=8cm,clip]{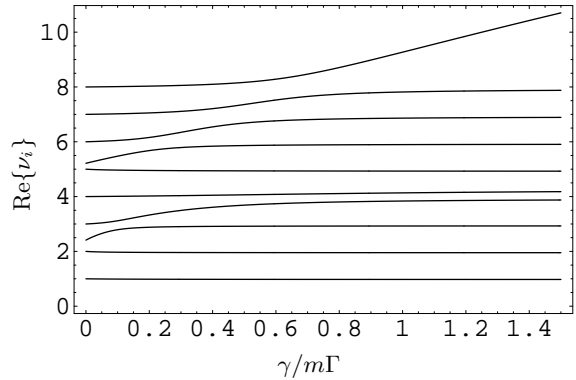}
 \caption[Bildunterschrift]{\label{EWswpkt}\sl\small
 Bath-coupling to the center of mass: eigenfrequencies $\nu$ of eq. \eqref{nuswpkt} for a finite model with $N+2=10$ oscillators versus the
 system-bath coupling strength $\gamma$ (in units of $m\Gamma$).
 Parameters: $\omega_a=2$, $\omega_b=5$, $m=m_{a,b}=1$, $D=2$ and $\Delta=1$, $\Gamma=N\Delta=8$.}\end{center}
\end{figure}

A more adequate model is the bath-coupling attached to the center of
mass $R$ of the system $H_{S'}$. The system-bath interaction could
then be described by the Hamiltonian:
\begin{equation}H_{S'E}=\sum_{i=1}^N\left[-c_ix_{i}R+\frac{c_i^2}{2m_{i}\omega_{i}^2}R^2\right].\end{equation}
This coupling leads to the following eigenvalue equation of the
total system $H_{\rm tot}$:
\begin{eqnarray}\label{nuswpkt}\nu^2-\frac{1}{M}(m_a\omega_a^2+m_b\omega_b^2)=\sum_{i=1}^N\frac{c_i^2}{Mm_{i}(\nu^2-\omega_{i}^2)}\frac{\nu^2}{\omega_{i}^2}-\nonumber\\
-\frac{\mu(\omega_a^2-\omega_b^2)^2}{m_b\omega_a^2+m_a\omega_b^2+\frac{M}{\mu}D-M\nu^2}\qquad.\end{eqnarray}
In Fig. \ref{EWswpkt} one can recognize that the lowest eigenvalue
is only slightly reduced by increasing the coupling strength and
remains real for all $\gamma$-values.\\
Since this interaction term assures the positive definiteness of the
total Hamiltonian we will use this system-bath coupling in the
further examination. Additionally this coupling allows us to
transform the system easily to normal coordinates which simplifies
the analysis in the case of identical oscillators.
\subsection{Langevin equation of two coupled Brownian oscillators}
By transforming the Hamiltonian $H_S$ onto coordinates for the
center of mass $R=1/M(m_aq_a+m_bq_b)$ and the relative coordinate
$x=(q_a-q_b)$ (with total mass $M$ and reduced mass $\mu$) and
eliminating the bath variables, the following system of coupled
equations for the Heisenberg operators $x$ and $R$ can be written
down:
\begin{eqnarray}\label{xgekopp} \ddot x &=& -\Omega_{x}^2 x(t)-(\omega_a^2-\omega_b^2)R(t)\\
\label{Rgekopp}\ddot R &=& -\Omega_{R}^2
R(t)-\frac{1}{M}\int_0^t\gamma(t-t')\dot
R(t)dt'-\gamma(t)R(0)-\nonumber\\&&-\frac{\mu}{M}(\omega_a^2-\omega_b^2)x(t)+\eta(t)\end{eqnarray}
with the frequencies
\begin{eqnarray}\Omega_x^2&=&\frac{1}{M}(m_b\omega_a^2+m_a\omega_b^2+\frac{M}{\mu}D)\\
\Omega_R^2&=&\frac{1}{M}(m_a\omega_a^2+m_b\omega_b^2)\end{eqnarray}
and damping term $\gamma(t)$ and noise term $\eta(t)$ as defined in
section II.A. Solving eq. \eqref{xgekopp} as an inhomogeneous
differential equation and inserting the solution into
\eqref{Rgekopp} gives a Langevin equation for the center of mass $R$
with new damping term $\tilde \gamma (t)$ and new noise term $\tilde
\eta(t)$:
\begin{equation}\label{Rlangevin2} M\ddot R + \frac{d\tilde V(R)}{dR} + \int_0^t{dt'\tilde{\gamma}(t-t') \dot R(t')} = \tilde{\eta}(t) - R(0)\tilde{\gamma}(t)\end{equation}
which describes the motion of $R$ in the effective potential $\tilde
V(R)=1/2M\tilde\Omega_R^2 R^2$ with frequency
\begin{equation}\qquad\label{tildeomegaR} \tilde \Omega_R^2=\Omega_R^2-\frac{\mu(\omega_a^2-\omega_b^2)^2}{M\Omega_x^2}\end{equation}
influenced by the damping
\begin{equation}\label{gammaR}\tilde{\gamma}(t-t')=\gamma\Gamma
e^{-\Gamma|t-t'|}+\mu\frac{(\omega_a^2-\omega_b^2)^2}{\Omega_x^2}\cos\Omega_x(t-t')\end{equation}
and the stochastic force
\begin{eqnarray}\tilde{\eta}(t)=\sum_{i=1}^N{c_i\left[x_i(0)\cos(\omega_it)+\frac{p_i(0)}{m_i\omega_i}\sin(\omega_it)\right]}+\nonumber\\+\mu(\omega_b^2
-\omega_a^2)\left[x(0)\cos(\Omega_xt)+\frac{p_x(0)}{M\Omega_x}\sin(\Omega_xt)\right].\quad\end{eqnarray}
In the case of identical oscillators the equations \eqref{xgekopp}
and \eqref{Rgekopp} are decoupled. The relative coordinate performs
a harmonic oscillation and for $R$ the Langevin equation of a
Brownian particle with mass $M$ and oscillator frequency $\Omega_R$
is received,
\begin{equation}
\label{Rlangevin3} M\ddot R + M \Omega_R^2 R +
\int_0^t{dt'\gamma(t-t') \dot R(t')} = \eta(t)-R(0)\gamma(t),
\end{equation}
which is equivalent to quantum Langevin equation \eqref{langevinBM}
in the first part of this paper.
\subsection{Stationary state}
\begin{figure}[ht]
\begin{center}
\psfrag{x1}[c]{$kT/\hbar\tilde\Omega_R$}\psfrag{x2}[c]{$\langle
R^2\rangle_{\rm cl}$, $\mbox{}\langle R^2\rangle_{\rm th}$,
$\mbox{}\langle R^2\rangle$}
 \includegraphics[width=8cm,clip]{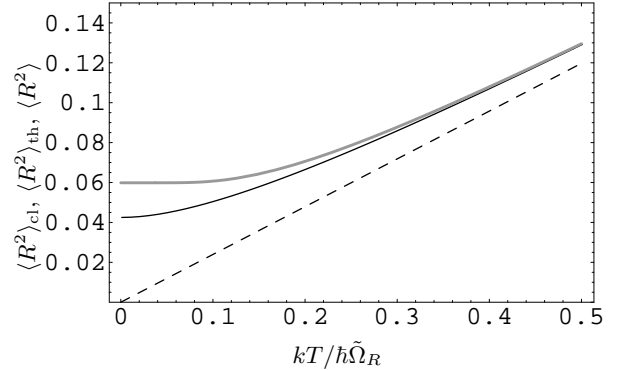}
 \caption[Bildunterschrift]{\label{RRvgl}\sl\small
  Temperature dependence of $\langle R^2\rangle$ \eqref{RReq} for parameter values $\omega_a=2$, $\omega_b=3\omega_a$,
   $m_a=m_b=1$, $D=M\Omega_R^2$, $\Gamma=10$ and $\gamma=5M\tilde\Omega_R^2/\Gamma$,  (dark line).
  For increasing temperature $\langle R^2\rangle$ converges to the limit cases $\langle R^2\rangle_{\rm th}$ (grey line) and $\langle R^2\rangle_{\rm cl}$ (dashed line) as given by eq. \eqref{RRbosevgl} and \eqref{RRclvgl}. } \end{center}
\end{figure}

In order to calculate the stationary correlations for the general
case we again apply the quantum fluctuation-dissipation-theorem.\\
From eq. \eqref{Rlangevin2} one obtains the dynamical susceptibility
$\tilde
\chi''_R(\omega)=[M\tilde\Omega_R^2-M\omega^2-i\omega\tilde\gamma(\omega)]^{-1}$
and can express the variance of the center of mass by
\begin{eqnarray}\label{RReq}\langle R^2\rangle&=&\frac{\hbar}{2\pi}\int_{-\infty}^{\infty}d\omega \tilde\chi_R''(\omega)\coth(\frac{1}{2} \beta \hbar
\omega)=\\&=&\frac{\hbar}{2\pi}\int_{-\infty}^{\infty}d\omega
\frac{\gamma \Gamma^2 \omega}{\xi^2\Gamma^2+(\xi
+\gamma\Gamma)^2\omega^2}\coth(\frac{1}{2}\beta\hbar\omega)\nonumber\end{eqnarray}
where
\begin{equation}\xi=M\Omega_R^2-M\omega^2-\mu\frac{(\omega_a^2-\omega_b^2)^2}{\Omega_x^2-\omega^2}.\end{equation}
Fig. \ref{RRvgl} shows the temperature dependence of the variance
\eqref{RReq} in comparison to the limiting cases:
\begin{eqnarray}\label{RRbosevgl}\langle R^2
\rangle_{\rm th}&=&\frac{\hbar}{2M\tilde\Omega_R}\coth\frac{1}{2}\beta\hbar\tilde\Omega_R\quad\mbox{for}\quad\gamma\to 0\\
\label{RRclvgl}\langle R^2 \rangle_{\rm
cl}&=&\frac{kT}{M\tilde\Omega_R^2}\qquad\mbox{for}\qquad T\gg
m\tilde\Omega_R^2.
\end{eqnarray}
\\

In the same way we can specify the variance of the center of mass
momentum $P_R$:
\begin{eqnarray}\label{PReq}\langle P_R^2\rangle&=&\frac{\hbar}{2\pi}M^2\int_{-\infty}^{\infty}d\omega \tilde\chi_R''(\omega) \omega^2 \coth(\frac{1}{2} \beta \hbar \omega)
=\\&=&\frac{\hbar}{2\pi}M^2\int_{-\infty}^{\infty}d\omega
\frac{\gamma \Gamma^2 \omega^3}{\xi^2\Gamma^2+(\xi
+\gamma\Gamma)^2\omega^2}\coth(\frac{1}{2}\beta\hbar\omega)\nonumber\end{eqnarray}
as well as the variance of the relative coordinate $x$
\begin{eqnarray}\label{xxeqrel}\langle x^2\rangle&=&\frac{1}{2\pi}\int_{-\infty}^{\infty}d\omega \hbar \coth(\frac{1}{2} \beta \hbar \omega)
\tilde\chi_x''(\omega)=\\&=&\frac{\hbar}{2\pi}\int_{-\infty}^{\infty}d\omega
\frac{(\omega_a^2-\omega_b^2)^2}{(\Omega_x^2-\omega^2)^2}\frac{\gamma
\Gamma^2
\omega\coth(\frac{1}{2}\beta\hbar\omega)}{\xi^2\Gamma^2+(\xi
+\gamma\Gamma)^2\omega^2}\nonumber\end{eqnarray} and the
corresponding momentum $p_x$
\begin{equation}
\label{pxeq}\langle
p_x^2\rangle=\frac{\hbar}{2\pi}\mu^2\int_{-\infty}^{\infty}d\omega
\frac{(\omega_a^2-\omega_b^2)^2}{(\Omega_x^2-\omega^2)^2}\frac{\gamma
\Gamma^2 \omega^3\coth(\frac{1}{2} \beta \hbar
\omega)}{\xi^2\Gamma^2+(\xi +\gamma\Gamma)^2\omega^2}.\end{equation}
The stationary correlation $\langle xR\rangle$ is obtained by
transforming on normal coordinates $y=\zeta x+R$, $z=x+\vartheta R$.
Because of $\langle yz\rangle=0$ in the stationary state one
receives
\begin{equation}\label{xReq}
\langle xR \rangle=\frac{\zeta\langle x^2 \rangle + \vartheta\langle
R^2\rangle}{1+\zeta\vartheta}
\end{equation}
\begin{eqnarray}
\mbox{where}\quad \zeta&=& - \frac{\mu (\omega_a^2-\omega_b^2)}{M\Omega_{-}^2-(m_b\omega_a^2+m_a\omega_b^2+M/\mu D)} \nonumber\\
\vartheta&=& -
\frac{M(\omega_a^2-\omega_b^2)}{M\Omega_{+}^2-(m_a\omega_a^2+m_b\omega_b^2)}\end{eqnarray}
and $\Omega_{\pm}$ are the normal frequency of system \eqref{Hs}
with $H_{AB}$ given by \eqref{Hac3}:\begin{eqnarray}\label{omegaS3}
{\Omega_{\pm}}^2&=&\frac{1}{2}\left(\omega_a^2+\omega_b^2+\frac{D}{\mu}\right)
\pm\\
&&\frac{1}{2}\sqrt{\left(\omega_a^2+\omega_b^2+\frac{D}{\mu}\right)^2-4D\left(\frac{\omega_a^2}{m_b}+\frac{\omega_b^2}{m_a}\right)-4\omega_a^2\omega_b^2}
\nonumber\end{eqnarray} Further correlations, such as $\langle
RP_R\rangle, \langle xp_x\rangle, \langle Rp_x\rangle,...$ are zero
in the stationary state.
\subsection{Thermodynamic of adiabatic changes}
The stationary Gaussian state of the subsystem is completely
characterized by the correlations \eqref{RReq},
\eqref{PReq}-\eqref{xReq}. The internal energy again is defined as
the stationary mean value of the systems Hamiltonian \eqref{Hs}:
\begin{eqnarray}
U_{S'}&=&\frac{1}{2M}\langle P_R^2\rangle+\frac{1}{2\mu}\langle
p_x^2\rangle+\frac{1}{2}\mu\Omega_x^2\langle
x^2\rangle+\nonumber\\&&+\frac{1}{2}M\Omega_R^2\langle
R^2\rangle+\mu(\omega_a^2-\omega_b^2)\langle xR\rangle.
\end{eqnarray}
In case of identical oscillators ($m_{a,b}=M/2$,
$\omega_{a,b}=\omega$) the internal energy $U_{S'}$ turns into
\begin{equation}
U_{S'}=\frac{1}{2M}\langle P_R^2\rangle+ \frac{1}{2}M\omega^2\langle
R^2\rangle +
\frac{1}{2}\hbar\Omega_{\mu}\coth(\frac{1}{2}\beta\hbar\Omega_{\mu}),
\end{equation}
where $\langle P_R^2\rangle$ and $\langle R^2\rangle$ are given by
the moments defined in equations \eqref{xxeq} and \eqref{ppeq} with
oscillator parameters $M$ and $\omega$. We can apply the weak
coupling limit to the (free) oscillation of the relative coordinate
so that the values $\langle x^2\rangle$ and $\langle p_x^2\rangle$
are determined by the quantum Gibbs distribution of an uncoupled
oscillator with mass $\mu$ and frequency
$\Omega_{\mu}=\sqrt{\omega^2+D/\mu}$. The motion of $R$ is described
by eq. \eqref{Rlangevin3} and leads to the stationary variances
given by \eqref{xxeq} and \eqref{ppeq}.\\

In this case of identical oscillators the von Neumann entropy
$S(\rho_{S'})$ of the system $H_{S'}$ can be expressed as sum of the
entropy of the center of mass coordinate $S_R$  and the entropy of
the relative coordinate $S_x$:
\begin{equation}S(\rho_{S'})=S_x+S_R,\end{equation}
where $S_R$ and $S_x$ are defined similar to equation \eqref{SvN}
with the phase space volumes \begin{eqnarray}v_x&=&\sqrt{\langle
x^2\rangle \langle p_x^2 \rangle}/\hbar\qquad\mbox{and}\\
v_R&=&\sqrt{\langle R^2\rangle \langle P_R^2
\rangle}/\hbar.\end{eqnarray}

 The exchange of heat $\delta Q$ and the change in entropy
$dS(\rho_{S'})$ are defined equivalent to eq. \eqref{dUS} and
\eqref{dSrho} by \begin{eqnarray}\delta
Q&=&\mbox{Tr}[H_{S'}d\rho_{S'}]\qquad\mbox{and}\\
dS&=&-\mbox{Tr}[d\rho_{S'}\ln\rho_{S'}].\end{eqnarray}

We now want to study deviations from the the Clausius inequality in
the case of identical oscillators. Regarding variation of the mass
$M$ the Clausius inequality reads:
\begin{equation}
\delta Q_M \leq TdS_M.
\end{equation}
With regard to an information theoretical viewpoint, we use again
the statistical entropy instead of the thermodynamic entropy $S_{\rm
th}$ for which the Clausius equality -- by
definition -- is fulfilled for quasi static processes.\\

\begin{figure}[h]
\psfrag{x1}[c]{$kT/\hbar\omega$}\psfrag{x2}[c]{$TdS_M$ and $\delta
Q_M$} \centering \subfigure[\sl \small \mbox{ } Coupling parameter
$D$ between the oscillators $A$ and $B$ is set to $D=M\omega^2$.]
{\includegraphics[width=8cm,clip]{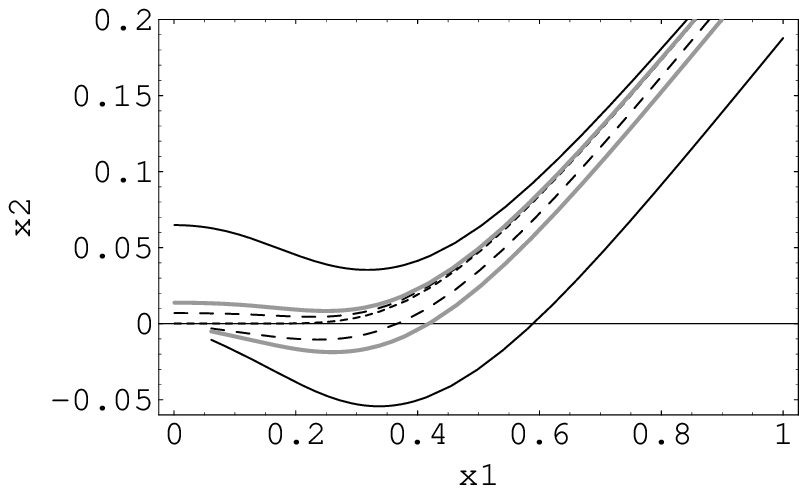}}\quad \subfigure[\sl
\small \mbox{ } Weak coupling between the oscillators $A$ and $B$
with parameter
$D=0.01M\omega^2$.]{\includegraphics[width=8cm,clip]{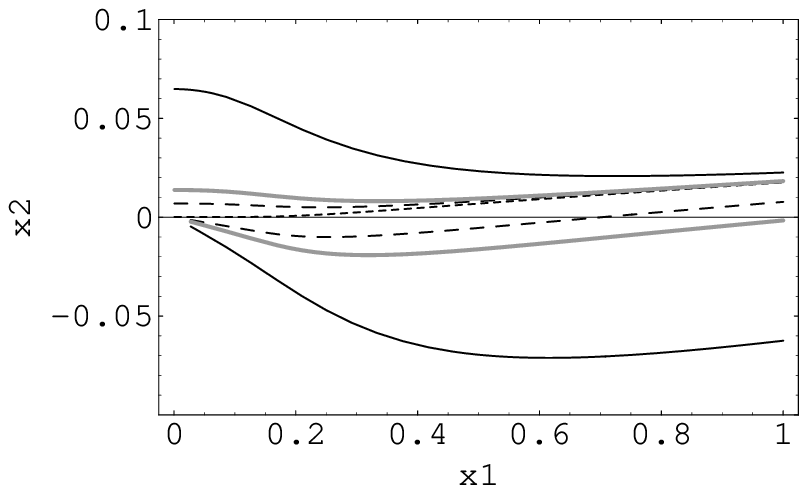}}
 \caption[Bildunterschrift]{\label{dQMdSMvsD}\sl\small
 Changes in heat $\delta Q_M$ and entropy-term $TdS_M$ versus bath temperature $T$
 for two different values of the oscillator coupling $D$. The system-bath coupling is
chosen as in fig. \ref{dQmdSmvsT}. From top to bottom: $\delta
Q_{M,\gamma=5}$, $\delta Q_{M,\gamma=1}$, $\delta Q_{M,\gamma=0.5}$,
$TdS_{M,\gamma=0.5}$, $TdS_{M,\gamma=1}$, $TdS_{M,\gamma=5}$. The
oscillator parameters are $\omega=\omega_{a,b}=1$ and
$M=2m_{a,b}=1$. The cutoff-frequency is set to $\Gamma=10$ and
$\hbar=1$. (Compare to fig. \ref{dQmdSmvsT}.)}
\end{figure}

In the weak-coupling case of a single harmonic oscillator with mass
$M$, entropy and heat are not affected by adiabatic variations of
the mass: $\delta Q_M=TdS_M=0$. This is different in the case of two
coupled oscillators weakly interacting with the bath, where the
additional coupling parameter $D$ leads to an increase of $\delta
Q_M$ and $TdS_M$ with rising temperature (narrow dashed curves in
fig. \ref{dQMdSMvsD}a and \ref{dQMdSMvsD}b). Nevertheless for
$\gamma \to 0$ the equality $\delta Q_M=TdS_M$ holds for all values
of $T$ and $D$. Furthermore, in this weak coupling approximation it
is $\delta Q_M\geq0$ and $dS_M\geq$ for all $T$ and $D$. \\

The impact of a non-zero system-bath-coupling $\gamma$ can be
studied from fig. \ref{dQMdSMvsD}a: the exchanged amount of heat
$\delta Q_M$ is increased at a given bath temperature $T$, whereas
the term $TdS_M$ is reduced. At low temperatures the product $TdS_M$
becomes negative, so that changes in heat and entropy have different
signs. This effect is the larger the stronger the
system-bath-coupling
$\gamma$ is. At high temperatures $kT\gg \hbar\omega$ both terms become equal.\\

If the coupling $D$ between the two oscillators is reduced ($D\to
0$), the system behaves like a single Brownian particle as could be
supposed by comparing the figures
\ref{dQMdSMvsD}b and \ref{dQmdSmvsT}.\\

\begin{figure}[t]
\psfrag{x1}[c]{$D/M\omega^2$}\psfrag{x2}[c]{$TdS_M$ and $\delta
Q_M$} \centering \subfigure[\sl \small \mbox{ } $\delta Q_M$ and
$TdS_M$ versus coupling D at bath temperature
$T=\hbar\omega/k$.]{\includegraphics[width=8cm,clip]{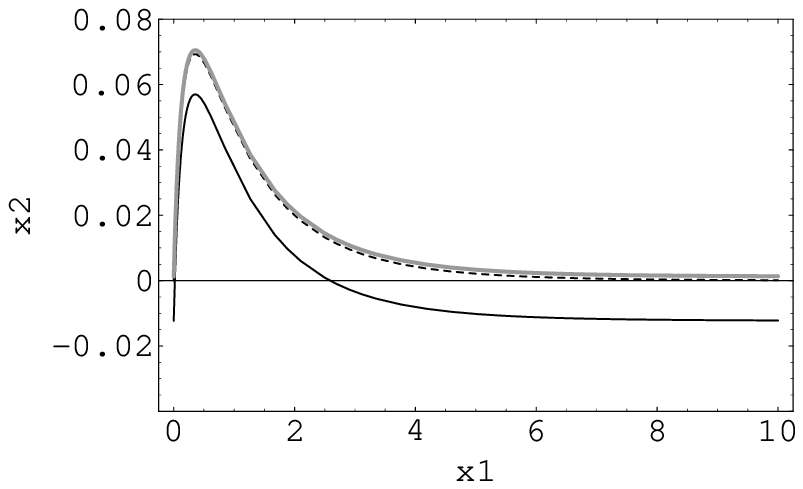}}\quad
\subfigure[\sl \small \mbox{ }$\delta Q_M$ and $TdS_M$ versus
coupling D at bath temperature
$T=0.5\hbar\omega/k$.]{\includegraphics[width=8cm,clip]{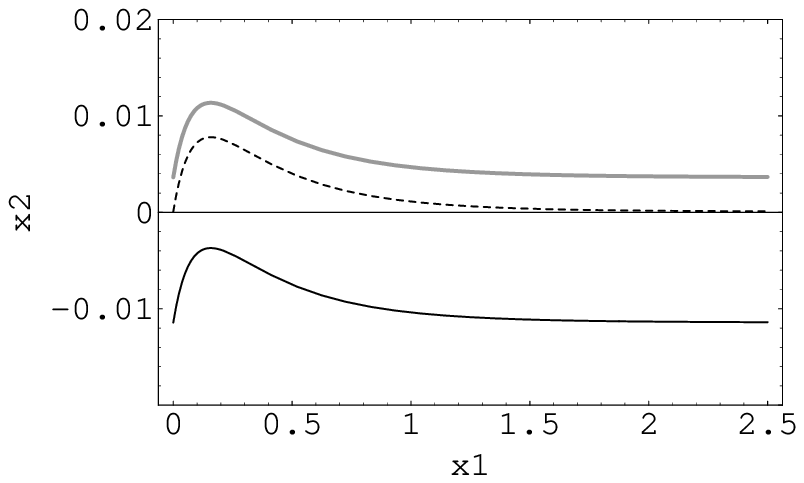}}
 \caption[adiabatic changes]{\label{DdSMdQMvsDT}\sl\small
 Changes in heat $\delta Q_M$ (grey line) and
entropy-term $TdS_M$ (dark line) versus coupling parameter $D$ at
two different bath temperature $T$
  in comparison to the case of $\gamma\to 0$ where $\delta Q_{M,\gamma=0}=TdS_{M,\gamma=0}$ (dashed line).
  Parameters are chosen as in fig. \ref{dQMdSMvsD}.}
\end{figure}

For $D \to \infty$ the system again behaves like a single Brownian
oscillator. This is shown by fig. \ref{DdSMdQMvsDT}a and fig.
\ref{DdSMdQMvsDT}b which give the $D$-dependence of $\delta Q_M$ and
$TdS_M$ for different temperatures $T$ and couplings $\gamma$. One
can recognize that the values of $\delta Q_M$ for $D\to\infty$ and
$D\to 0$ are equal (as well as the values for $TdS_M$) and
correspond with the results for a single Brownian oscillator.\\

Furthermore, fig. \ref{DdSMdQMvsDT}a shows that, depending on the
chosen parameter values of oscillators ($M$, $\omega$) and bath
($\gamma$, $\Gamma$, $T$), there may exist a range of $D$-values
where the terms $TdS_M$ and $\delta Q_M$ have equal signs. Fig.
\ref{DdSMdQMvsDT}b makes clear that at lower temperatures such a
region does not exist necessarily.\\
Summarizing the results of this section, we can state the
following:\\ Compared to the case of a single Brownian particle, an
open quantum system with internal degrees of freedom shows
additional effects. In our introduced model the direction of heat
and entropy flow due to mass variations depends on the coupling
strength between both the oscillators. Already at moderate
temperatures the flow of heat and entropy occurs in the same
direction, whereas in the model of single Brownian motion this is
reached only in the high temperature and weak coupling limit
respectively. Therefore varying the coupling parameter offers the
possibility to adjust the ratio of heat exchange and change in the
subsystems entropy.\\
Of course, the resulting changes in heat and entropy depend on the
chosen interaction between the oscillators. As pointed out at the
beginning of the section, this coupling had to be selected carefully
to ensure the positive definiteness of the total Hamiltonian.
Nevertheless this model provides a good example for studying the
influence of additional degrees of freedom.

\section{summary and conclusions}
We have discussed the statistical thermodynamics of quantum Brownian
motion of two coupled oscillators in the strong coupling quantum
regime and have compared the results to the case of a single
Brownian particle. In both cases quantum correlations between
subsystem and bath lead to deviations from the canonical equilibrium
thermodynamics. The quantum Langevin equation which has been derived
for a system of two coupled Brownian oscillators describes the
evolution of the Heisenberg operators. The stationary moments of
these operators characterize the reduced density matrix completely.
This density matrix contains all the accessible information about
the quantum state. The statistical entropy of this state is measured
by the von Neumann entropy. Reducing this entropy by quasi-static
parameter variations is equivalent to a decrease in the information
which is gained by a measurement. With regard to continuous variable
quantum computing we examined the relation between changes in the
subsystems statistical entropy and the exchange of heat with the
environment. We found that this relation deviates from the Clausius
(in)equality at low temperatures due to the existing correlations
between system and bath. Related results of former work
\cite{Nieuwenhuizen1, Nieuwenhuizen2, Nieuwenhuizen8} were
nevertheless put into perspective. Concerning quantum information
processing, the validity of the Landauer principle which is based on
the Clausius inequality but deals with the statistical entropy seems
indeed questionable -- at least for open quantum systems which are
non-weakly interaction with a low temperature environment. \\

{\bf Acknowledgement}

 One of the authors (C.H.) appreciates a fruitful
discussion with Peter H\"anggi which helped to clarify some of the
controversial points.

\bibliographystyle{apsrev}
\end{document}